\documentstyle[twoside,fleqn]{article}

\newcommand{\bee}{\begin{equation}}
\newcommand{\ene}{\end{equation}}

\title{Spectral geometry for strings and branes}

\author{{Dmitri V. Vassilevich}\\
{Institut f\"{u}r Theoretische Physik, Universit\"{a}t Leipzig,}\\
{Augustusplatz 10/11, D-04109 Leipzig, Germany}\\{and}\\
{V.~A.~Fock Institute of Physics, St.Petersburg University,}\\
{198904 St. Petersburg, Russia}%
\thanks{E.mail: Dmitri.Vassilevich@itp.uni-leipzig.de}}

\begin{document}

\maketitle
\begin{abstract}
I give a short guide into applications of the heat kernel technique
to the string/brane physics with an emphasis on the emerging boundary
value problems.
\end{abstract}

\section{Introduction}
The heat kernel technique is especially effective for evaluation
of the one-loop beta functions, quantum anomalies and low-energy
effective action. These three topics cover a large part of
calculations in the sigma model approach to string theory.
Local geometry of the closed strings is too simple to reveal
advantages of the heat kernel methods. For open strings 
power of these methods was recognized long ago (see e.g.
\cite{Alvarez:1983zi}). Now, boundary conditions
and boundary dynamics are of primary interest in string/brane
theory. 

In this note I give a short review of boundary value problems
which appear in various string and brane models, list known
properties of these problems and present some open questions.
We will mostly deal with partial differential operators of Laplace
type acting on smooth sections of a vector bundle. 
Any operator of this type can be uniquely represented as
\begin{equation}
D=-(\nabla^\mu \nabla_\mu + E(x)) \label{Lop}
\end{equation}
with an appropriate connection $\nabla$ and an endomorphism
(matrix-valued function) $E$. Usually, but not always, there is
a power law asymptotics of the heat kernel as $t\to +0$
\begin{equation}
{\rm Tr} (fe^{-tD}) \simeq \sum_{n=0} t^{(n-m)/2} a_n(f,D)\,, \label{asympt}
\end{equation}
where $m$ is dimension of the manifold, and the heat kernel
coefficients are $a_n$ can be represented through bulk and surface
integrals.

Actual calculations of the coefficients $a_n$ are not easy. However,
these calculation are not as hard as one can imagine looking at
(very) long resulting expressions. For manifolds with boundaries the
general idea of the most powerful (functorial) method can be found in
\cite{BG90} (with some minor corrections in \cite{Vassilevich:1995we}).
The heat kernel coefficients for Dirichlet, Robin and mixed boundary
conditions up to $a_5$ can be found in \cite{Branson:1999jz}.

This paper is organized as follows. The next section is devoted to open
strings, Dirichlet branes and some more general objects. Non-local
mixtures (bound states) of strings and branes are described in sec.~3.
In sec.~4 I comment on the heat kernel expansion for singular geometries
of the brane-world type. Sec.~5 contains concluding remarks.

\section{Open strings and Dirichlet branes}
If we adopt Euclidean signature on both the target space and the
world sheet ${\cal M}$ and neglect dimensional couplings, the action for the
open string sigma model takes the form:
\begin{eqnarray}
&&S= \int _{\cal M}d^{2}z\sqrt{h}
\left(\frac{1}{2} G_{\mu\nu} h^{ab}
\partial _{a}X^{\mu }\partial _{b}X^{\nu }\right.\nonumber \\
&&\qquad \left. +\frac{1}{2\sqrt{h}}\epsilon^{ab}B_{\mu\nu}
\partial _{a}X^{\mu }\partial _{b}X^{\nu } \right)
\nonumber \\
&&\qquad+\int _{\partial {\cal M}}d\tau \left(
 A_{\mu }\partial _{\tau }
X^{\mu }  
\right). \label{act}
\end{eqnarray}
Here we adopted the standard notations of \cite{GSW}.
Let us split the string coordinate $X=\bar X+\xi$ into background
part $\bar X$ and fluctuations $\xi$. Keeping the second order
terms in $\xi$ which are relevant for the one-loop corrections
we obtain:
\begin{eqnarray}
&&S_2=\frac 12 \int _{\cal M}d^{2}z\sqrt{h} \xi D \xi  \label{bterm} \\
&&\ +\frac 12 \int _{\partial {\cal M}}d\tau \xi \left(
\nabla_N +\frac 12 (\nabla_\tau \Gamma + \Gamma \nabla_\tau )
+{\cal S} \right)\xi.\nonumber
\end{eqnarray}
The operator $D$, the covariant derivative $\nabla$ and the 
matrix-valued functions $\Gamma$ and ${\cal S}$ depend on the
background fields $A(\bar X),\ B(\bar X),\ G(\bar X)$.
The boundary conditions should satisfy
two important requirements: i) the operator $D$ should be
formally self-adjoint (symmetric); ii) the boundary term
in (\ref{bterm}) should vanish. There are two obvious local
choices of such boundary conditions:
\begin{equation}
\left( \nabla_N +\frac 12 (\nabla_\tau \Gamma + \Gamma \nabla_\tau )
+{\cal S} \right)\xi \vert_{\partial{\cal M}}=0 \label{opstr}
\end{equation}
corresponding to open strings, and
\begin{equation}
\xi \vert_{\partial{\cal M}}=0 
\label{Dir}
\end{equation}
describing the Dirichlet branes. Generically, some of the components
of $\xi$ satisfy (\ref{opstr}) while the other satisfy (\ref{Dir}).

The boundary conditions (\ref{opstr}) contain both normal and
tangential derivatives. Therefore, there are crucial differences
from ordinary Neumann or Robin boundary value problem.
In some literature the boundary conditions of the type (\ref{opstr})
are called oblique. Study of corresponding boundary value problem
has been initiated by Grubb \cite{Grubb} and then continued in
\cite{GS,McO,DK,AE}. Let us briefly summarize the relevant results.
On manifolds with boundary to ensure normal properties of the spectrum
of an operator of Laplace type (as e.g. no more than finite number
of negative modes) the boundary value problem must satisfy the so-called
strong ellipticity condition. In the present context it requires
that norm $\Gamma$ is not too large (meaning that the gauge field
strength is less than its critical value)\footnote{Note, 
that Dirichlet and Robin
boundary value problems are always strongly elliptic.}.
If the boundary value problem is strongly elliptic, there is an
asymptotic expansion (\ref{asympt}). Coefficients $a_n$ with $n\le 3$
for oblique boundary conditions have been calculated in \cite{McO,DK,AE}.

The analysis \cite{Osborn91} of the conformal invariance of open strings
performed with the help of the heat kernel expansion confirms,
in general, the earlier results \cite{Callan}. However, some additional
terms in the conformal variations appear \cite{Osborn91} (see also
\cite{KVunp}). Interpretation of these terms is still unclear.

In \cite{Kummer:2000ae} we argued that multiplicative renormalizability
of the open string sigma model suggests introduction in the action
(\ref{act}) a new surface coupling:
\begin{equation}
-\int_{\partial {\cal M}} V_\mu (X)\partial_N X^\mu \,. 
\label{Vcoupl}
\end{equation}
Such coupling in fact appeared already in \cite{Vc}.
Acting as before, we arrive at the following boundary conditions:
\begin{equation}
(-(1+\Lambda )\nabla_N \xi +L\xi )\vert_{\partial {\cal M}}=0\,,
\label{Vbc}
\end{equation}
where the operator $L$ contains tangential derivatives only, the matrix
$\Lambda$ depends on $V$: $\Lambda_{\mu\nu}=D_\mu V_\nu$. Since the
operator $L$ is defined by the quadratic form of the boundary action,
it is fixed up to an anti-hermitian part (which vanishes being sandwiched
between two $\xi$). This ambiguity is to be used to make the operator
$(1+\Lambda )^{-1}L$ a hermitian operator on the boundary (to ensure
hermiticity of the volume operator $D$). Solution for $L$ has a rather
complicated form. Consequently, the heat kernel calculations also become
complicated. Presently, the heat kernel coefficients are known in
few first orders of $\Lambda$ or, equivalently, of $V$ 
\cite{Kummer:2000ae}.

Increasing complexity of the calculations was not our motivation.
The boundary conditions (\ref{Vbc}) possess a very interesting 
qualitative property. They interpolate between the open string boundary
conditions $\Lambda =0$ and the D-brane boundary conditions $\Lambda =-1$.
This suggest a promising mechanism of spontaneous Dirichlet brane
creation. Actual realization of this mechanism requires more effective
methods of calculation of the beta-functions or of the heat kernel
expansion.  
\section{Spectral branes}
As has been already mentioned in the previous section, a consistent 
string sigma model is described by a mixture of the boundary conditions
(\ref{opstr}) and (\ref{Dir}). For the Dirichlet branes such a mixture
is prepared in a particularly simple way: it is enough to say that
the $p+1$ components satisfy (\ref{opstr}) while the rest of the
components obey (\ref{Dir}). This picture can be easily visualized,
but it is not unique. Projector on the Dirichlet components may be
as well a complicated non-local operator. An example of such non-local
construction was given in \cite{Vassilevich:2001at}. It uses the famous
spectral boundary conditions of Atiyah, Patodi and Singer \cite{Atiyah:1980jh}.
Such boundary conditions are defined with respect to a Dirac operator.
We may take a ``square root'' $P$ of the Laplace operator $D$ appearing
in the quadratic form of the action  (\ref{bterm}) defined as
$P^\dag P=D$ (if such a ``square root'' exists). To this end we have
to find a representation of the Clifford algebra in two dimensions with
the target space indices. Effectively, one has to decompose the target
space into a direct sum of two-dimensional subspaces
and define standard $\gamma$-matrices on each subspace. Consider the
simplest case when $D$ is just the free Laplacian. Then 
$P=\gamma^a\partial_a$. Define 
${\cal H}=\gamma_\sigma \gamma^\tau \partial_\tau$, where $\tau$ is a
geodesic coordinate on the boundary, $\sigma$ is a normal coordinate.
Now, we may impose the boundary conditions (\ref{opstr}) and (\ref{Dir})
independently on each eigenmode of ${\cal H}$. For example, we may say
that the modes with positive eigenvalues of ${\cal H}$ satisfy (\ref{opstr}),
and the rest obey (\ref{Dir}). This is the essence of the $S$-brane
construction. This object has a lot of interesting properties 
\cite{Vassilevich:2001at}. Most remarkable is the behavior of the
$S$-branes under the T-duality transformation (see, e.g., \cite{Buscher:1987sk}
for a path integral derivation of the T-duality). $S$-branes may be mapped
to themselves, to open strings and to $D$-branes. Therefore,
$S$-branes play a rather special role in string physics. 
However, many points still have to be clarified. This includes
classification of the Laplace operators which are representable through
the Dirac operators. Heat trace asymptotics for these rather complicated
boundary value problem have not been calculated so far.

\section{Brane-world}
The brane-world scenario \cite{b-w}
suggests a rather unusual type of the
``boundary value'' problem. It involves a singular surface $\Sigma$
such that only the leading symbol of the operator $D$ (i.e., the
Riemann metric) is assumed to be continuous across $\Sigma$.
All other geometric quantities including derivatives of the
metric can jump on $\Sigma$. One has to assume some matching
conditions on the limiting values of the field and of it's normal
derivative. Until very recent time the heat kernel expansion
for that kind of problems escaped systematic study.
As far is the standard brane world scenario is concerned,
the heat kernel asymptotics can be found in
\cite{Gilkey:2001hh,singular}. However, calculations for the
most general matching conditions \cite{AGHH88} are yet to be performed.

\section{Conclusions}
We have described several boundary value problems which appear in
the string or brane context. As usually, one needs the $\beta$-functions
and the effective action in the form of an expansion in the background
fields. These problems can be solved by the heat kernel methods.

A non-standard application of the heat kernel technique motivated
by strings is duality relations between functional determinants
(see e.g. \cite{dual}). The simplest example of such relations
is transformation of the effective action $W[\phi ]=-\frac 12
\log\det (-e^{\phi} \nabla_\mu e^{-2\phi} \nabla^\mu e^\phi )$
under change of the sign in front of the dilaton field $\phi$.
In two dimensions $W[\phi ]-W[-\phi ]$ is a local functional of
$\phi$ which can be expressed through the heat kernel coefficients.
There are also higher dimensional generalizations of this relation.
The dilaton $\phi$ can be replaced by a matrix-valued function.
Apart from strings the dilaton interaction of the type described
above appears also from the spherical reduction of higher dimensional
theories \cite{sred}.

Rather surprisingly, despite constant attention to this field
of research, many problems are still left open. String models
provide large number of applications for the spectral geometry methods
which are the key subject of the present Conference. 
\section*{Acknowledgements}
I am grateful to M.~Bordag, P.~B.~Gilkey, K.~Kirsten, W.~Kummer
and A.~Zelnikov
for collaboration. This work has been supported in part by the
DFG project Bo 1112/11-1.

\end{document}